\documentstyle[emulateapj]{article}

\def\submitted#1{\gdef\@submitted{#1}}
\def\subtitle{
  \vspace*{-12mm}
  \noindent
  {\scriptsize {\sc \@submitted}}
}
\let\slugcomment\submitted   
\submitted{Draft version \today}

\def\lesssim{\mathrel{\hbox{\rlap{\hbox{\lower4pt\hbox{$\sim$}}}\hbox{$<$}}}}
\def\gtrsim{\mathrel{\hbox{\rlap{\hbox{\lower4pt\hbox{$\sim$}}}\hbox{$>$}}}}

\def\arcdeg{\hbox{$^\circ$}}

\newcommand{\mamo}[1]{\mbox{$#1$}}
\newcommand{\unit}[1]{\ifmmode \:\mbox{\rm #1}\else \mbox{#1}\fi}
\renewcommand{\sb}[1]{_{\rm #1}}
\newcommand{\mone}{\mamo{^{-1}}}

\newcommand{\kms}{\unit{km~s\mone}}

\newcommand{\mpc}{\unit{Mpc}}

\newcommand{\hmpc}{\mamo{h\mone}\mpc}

\newcommand{\lb}[2]{\mamo{l = #1\arcdeg}, \mamo{b = #2\arcdeg}}

\newcommand{\eqref}[1]{equation~(\ref{eq:#1})}

\slugcomment{To appear PASP, November 99\\[3ex]
{\large \it \bf Conference Highlights \footnote{
Conference was held in Victoria, BC, Canada in July
1999. Proceedings will be edited by St\'{e}phane Courteau, Michael
Strauss \& Jeffrey Willick and published in the {\it ASP Conference
Series}.}}
}

\lefthead{M. J. HUDSON}
\righthead{Cosmic Flows}

\begin{document}

\title{Cosmic Flows: Towards an Understanding of the
Large-scale Structure in the Universe}

\author{
Michael J. Hudson
}
\affil{
  Department of Physics \& Astronomy, University of Victoria,
  P.O. Box 3055, Victoria, B.C. V8W 3P6, Canada.
  E-mail: hudson@uvastro.phys.uvic.ca}

\section*{}
This workshop was the first meeting devoted exclusively to ``Cosmic
Flows'' since the conference held at the Institut d'Astrophysique de
Paris exactly 6 years earlier.  The major issues addressed in Victoria
were summarized by Strauss in his introductory review:
\begin{enumerate}
\item 
On the largest scales, have we identified a volume which is at rest
with respect to the Cosmic Microwave Background (CMB)?
\item What can be learned from the ``coldness'' of the flow field on
small scales?
\item
Can we agree on a value of $\beta \equiv \Omega^{0.6}/b$?  If not, is
it because the relationship between mass and light (``biasing'') is
not a simple linear function?  How can peculiar velocity studies be
used to constrain biasing?
\item What are the next generation of peculiar velocity surveys?
\end{enumerate}

\section{Seeking convergence to the CMB frame}
One of the long-standing goals of observational cosmology has been to
identify a fair sample of the Universe. Within the context of cosmic
flows, this amounts to identifying the depth at which the local
Universe is at rest with respect to the CMB, and thus contains all of
the major attractors responsible for the 600 \kms\ motion of the Local
Group.  In practice, this can be done by selecting a sample of
galaxies for which distances are known and determining their net bulk
peculiar velocity in the CMB frame.

Coming into this workshop, the situation to a depth of $\lesssim 50
\hmpc$ was fairly clear: existing peculiar velocity catalogs indicated
that the CMB-frame bulk motion of this volume is $\sim 300\pm100
\kms$. A tidal analysis (Dekel)\footnote{Throughout this summary only
the individual who presented results in Victoria is referenced.}
indicates that most of this motion is due to masses beyond $\lesssim
50 \hmpc$.  However, newly-presented observational results on these
scales favor the lower end of this range, and thus more local sources.
Tonry \& Dressler obtained precise distances to nearby ($\lesssim 30$
\hmpc) galaxies with the Surface Brightness Fluctuation (SBF)
technique.  They modeled the flow field with Virgo and a Great
Attractor (GA) and found a GA infall velocity of $289\pm137$ \kms\ at
the position of the LG. (This is about half of that originally claimed
by the 7 Samurai.)  Furthermore, Tonry \& Dressler claim only a small
($\lesssim 200 \kms$) residual motion arising from sources beyond
their volume.  The ENEAR project, a Fundamental Plane (FP)-based
`son-of 7-Samurai' survey, finds somewhat intermediate results: a bulk
flow of $343\pm53$ \kms\ at a depth of $\sim 50 \hmpc$ (Wegner). The
`Shellflow' project (Courteau) using the Tully-Fisher (TF) distance
indicator, found that galaxies in an all-sky shell centered on $55$
\hmpc\ have a low bulk motion of $80 \pm 150\kms$.  Taken together
these new results suggest that most of the mass responsible for the
motion of the LG lies within 60 \hmpc, and can presumably be
identified with the GA.  Woudt showed results from galaxy redshift
surveys in the Zone of Avoidance which indicate an excess of galaxies
in the GA region centered on Abell 3627 at \lb{325}{-7}. A3627 appears
to have the richness of Coma, and its position coincides with the peak
of the GA as inferred from POTENT reconstructions.

The observational situation on larger ($>60 \hmpc$) scales remains
complicated.  At the '93 Paris meeting, the controversial results of
Lauer \& Postman (LP) were dissected.  In Victoria, a number of recent
results probing similar scales were presented; these were based mainly
on sparse samples of galaxy clusters.  The FP-based SMAC sample
(Smith) and the TF-based LP10k sample (Willick) both find a large bulk
flow, $\sim 600\pm200 \kms$ (but in a direction different from LP),
whereas the TF-based SCI+II (Giovanelli \& Dale) sample yields a very
low bulk motion ($\sim 100 \pm 100$ \kms).  The nearby SNIa (Riess)
give an intermediate result $\sim 270 \kms$, in good directional
agreement with SMAC/LP10k. The EFAR survey (Colless) finds low
peculiar motions, but is restricted to two regions in the sky which do
not sample well the directions of the SMAC/LP10k flows. One might
wonder if these different groups measure the same peculiar velocities
for common clusters.  The answer is a qualified yes: formally -- given
the large errors -- the $\chi^2$ is acceptable, but there may be
a trend for some surveys have systematically lower peculiar
velocities than others (Lucey).  It is important to note that what is
quoted is the bulk flow of a given sparse sample and not that of a
volume. I argued that given the sampling and the presence of internal
flows --- which act like an extra noise term --- all of these surveys
(with the exception of LP) are consistent with each other and with a
mean large-scale bulk flow of $\sim 300-400 \kms$.  This is somewhat
larger than the bulk motions obtained on smaller scales, but the
difference may not be statistically significant. Ultimately, larger
and denser samples will be needed to determine the contribution from
the largest scales.

Peculiar velocity data can be used to place direct constraints on the
mass power spectrum on large scales.  Zehavi showed that peculiar
velocity data sets gave the constraint $\Omega^{0.6} \sigma_8 \sim
0.8$, somewhat higher than the cluster normalization.

In order to probe the mass power spectrum on small scales, a statistic
such as the r.m.s. velocity dispersion of galaxies around the
large-scale flow can be used.  Baker showed that the 1D single-object
weighted velocity dispersion of Las Campanas redshift survey (LCRS)
galaxies is $\sim 125 \kms$, which favors $\Omega^{0.5}/b \sim
0.3$--$0.6$.  This dispersion is similar to the velocity scatter of
SBF galaxies around the best fitting flow models (180-200 \kms,
Blakeslee; Tonry \& Dressler). Recent theoretical developments on the
pairwise infall velocity (Juskiewicz) yield constraints on $\Omega$
and $\sigma_8$ which are consistent with those obtained from the
abundance of rich clusters. It remains to be determined, however, how
much the observed small-scale velocity dispersion depends on the
biasing of galaxies with respect to dark matter.

\section{Mass, Light and ``Biasing''}
On large ($> 10 \hmpc$) scales, there appears to be good agreement
between mass and light.  Dekel showed new POTENT and Weiner Filter
reconstructions of the local mass-density field --- based on the SBF
distances of Tonry et al.\ --- which were in good qualitative
agreement with Tully's galaxy maps of the nearby Universe.  When
fluctuations in the mass and galaxy densities are much less than the
mean density, as they are on large scales, it is reasonable to
parameterize the ratio of fluctuations by a single ``biasing''
parameter, $b$.  A comparison of mass and light galaxies then yields a
value of $\beta \equiv \Omega^{0.6}/b$.

There was considerable discussion of $\beta$ estimates derived from
different methods of comparing mass and light.  Comparisons of
predicted and observed peculiar velocities consistently yield low
values $\beta\sb{IRAS} \sim 0.45$ (e.g. Blakeslee; Riess). On the
other hand, the comparison of the galaxy density with mass density
reconstructed from POTENT yields $\beta\sb{IRAS} \sim 0.9$%
\footnote{
The power spectrum analysis of Zehavi yields even higher values:
corresponding to $\beta\sb{IRAS} \sim 1$, if $b\sb{IRAS}\sigma_8 \sim
0.75$.
}. Both methods have been extensively tested with N-body simulations and
mock catalogs (e.g. Kolatt's POTENT error analysis).  It is possible
that this discrepancy is due to the different scales probed:
velocity-velocity comparisons are performed on small smoothing scales,
whereas the density comparison is smoothed on larger scales. If this is
coupled with a scale-dependent biasing factor, as indicated by some
semi-analytic models of galaxy formation, then it might account for
the different results.


Indeed, on small (few \hmpc) scales, the true relation between mass
and light is complicated (Dekel); it is likely to be non-linear, a
function of scale, galaxy type, luminosity and surface brightness and
may depend on parameters other than the local mass density.  While the
clustering of dark matter haloes is now quite well understood through
N-body simulations (Frenk; Lake; Klypin) and analytic work (Sheth),
the star formation history of a given halo is more uncertain.
Pioneering attempts to quantify and explore non-trivial biasing using
LCRS and IRAS PSC redshift survey (PSCz) data were presented by
Blanton and Narayanan, respectively.  Shaya argued that the best fit
flow model for the nearby ($\lesssim 30 \hmpc$) Universe was one in
which early-type galaxies in dense cores were assigned a mass-to-light
ratio of $750$ whereas spirals in the field have $M/L \sim 150$.
Saunders commented that, in the PSCz catalog, the Hercules and Shapley
superclusters are of comparable overdensity, whereas if one counts
rich clusters, the latter is far more overdense.  Indeed, preliminary
POTENT reconstructions of SMAC+SNIa data indicate a large mass near
Shapley (Dekel).  It is clear, however, that much more work needs to
be done in order to understand the sensitivity of various statistics
to simple modifications of the linear biasing scheme.

\section{The Future}
Can we improve on current data sets?  The 2 Micron All-Sky Survey will
obtain redshifts for $\sim 10^5$ galaxies (Huchra) and should provide
superlative all-sky galaxy maps for comparison with peculiar velocity
surveys.  This effort will be complemented by the 6dF survey in
Australia which will also obtain velocity dispersions and thus
peculiar velocities for all E/S0 galaxies in the South (Mamon).
Furthermore, several of the TF/FP teams present in Victoria have plans
to extend their cluster or field-based studies.

Other methods also show promise. At present, SNIa searches are turning
up $\sim 20$ local ($cz < 30000 \kms$) per year, but this should
increase with new searches underway (Aldering).
The advent of 8m-class telescopes with good seeing will allow the SBF
technique to be pushed to $\sim 6000 \kms$, allowing the definitive
study of the local flow field.  Finally there remains the possibility
of using the X-ray emission from clusters as a distance indicator at
low redshift, and the kinematic Sunyaev-Zel'dovich effect to probe
cluster peculiar velocities at high redshift.  When these new data are
compared with sophisticated mock catalogs based on semi-analytic
galaxy formation models (Dekel), important insights will be gained
regarding the relationship between dark and luminous matter.

Overall, the most rewarding aspects of this workshop were the
thoughtful questions and very frank responses, both to individual
presentations and in the panel-led discussions. These discussions
raised more questions than they answered, as they should in any
fruitful field. St\'{e}phane Courteau is to be congratulated for
organizing a very successful and stimulating meeting.

\end{document}